\documentclass[preprint]{acmart}

\usepackage[utf8]{inputenc}

\usepackage{graphicx}
\usepackage{verbatim}

\newcommand{\xfdb}{XamForumDB }
\newcommand{\xfdbse}{XamForumDB}

\newcommand{\ent}[1]{{\bf{#1}}}
\usepackage{paralist}
\usepackage[export]{adjustbox}
\usepackage{numprint}
\npthousandsep{,}

\title{\xfdbse: a dataset for studying Q\&A about cross-platform mobile applications development.}

\author{
Matias Martinez, Sylvain Lecomte\\
	University of Valenciennes, France\\
first name.last name@univ-valenciennes.fr
}

\begin{document}


\begin{abstract}


Android and iSO are the two mobile platforms present in almost all smartphones build during last years.
Developing an application that targets both platforms is a challenge. A traditional way is to build two different apps, one in Java for Android, the other in Objective-C for iOS.
Xamarin is a framework for developing  Android and iOS apps which allows developers to share most of the application code across multiple implementations of the app, each for a specific platform.
In this paper, we present \xfdbse, a database that stores discussions, questions and answers extracted from the Xamarin forum.
We envision research community could use it for studying, for instance, the problematic of developing such kind of applications.
\end{abstract}

\maketitle

\section{Introduction}

Nowadays, there are billions of smartphone devices around the world which run Android or iOS operating systems.
A \emph{cross-platform mobile application} is an application that targets more than one mobile platform. For example, most of the famous social apps are available for both Android and iOS platforms.

A traditional approach for developing this kind of apps is to build a \emph{native} application for each platform.
Here, each native apps is build using a specific programming language (e.g., Java for Android, Objective-C vs Swift for iOS), SDK and IDE (e.g., Android Studio, XCode for iOS).
Unfortunately, this development process increases costs of development and maintenance phases. For example, the mobile development team must be competent to each target mobile platform. 

During last years, researches and industry sectors have both focused on proposing solutions for developing mobile applications with the goal of overcoming the mentioned problematic. 
One of those solutions is Xamarin\footnote{\url{https://www.xamarin.com/}}, a  framework to build native Android, iOS and Windows development in C\# programming language.

The advantage of Xamarin platforms is that developers re-use their existing C\# code, and share significant code across device platforms. For example, the app Evolve\footnote{\url{https://github.com/xamarinhq/app-evolve}} for managing conferences' events, was developed using Xamarin and is around 15,000 lines of code: the iOS version contains 93\% shared code i.e., C\# (the remain 7\% iOS specific code) and the Android version contains 90\% shared code\footnote{https://goo.gl/eqj1ml}.

Xamarin has an official online forum\footnote{https://forums.xamarin.com/}, which has multiple purpose: 
\begin{inparaenum}[a)]
\item to be a platform for communicating announcements (events, releases, jobs related with the platform) 
and 
\item to be a Q\&A site, where Xamarin users (such as cross-platforms developers) post discussions, questions and share answers. 
Also, the developers of the Xamarin platforms (i.e., members of the company that develop it) participate on the forum.
\end{inparaenum}

%

However, to the best of our knowledge, no work studies the quality of cross-platforms mobile apps, nor analyzes this source of information
for understanding the development process 
of mobile apps using a cross-platform development tool such as Xamarin.

In this paper, we present \xfdbse, a database that contains all discussions, questions and answers from the Xamarin forum. Moreover, it stores the number of views of each post, number of `likes', the accepted answers, and profile of the forum's user.

We consider that \xfdb gives to the software engineering research community the possibility to study the development and maintenance phases of cross-platform mobile applications, and to propose new approaches for improving the quality of mobile applications.
%
The \xfdb is publicly available at \url{https://goo.gl/A1eh03}

\section{Describing Forum Xamarin}


The Xamarin forums site is a online web platform where mobile developers  post questions or start a discussion about the Xamarin framework and its ecosystem.
Moreover, it is a communication channel between the developers of the framework (a.k.a. Xamarin Team) and users of it (i.e., the developers of mobile apps that use Xamarin as development framework).
For instance, new version of the framework or particular components are announced in the forum.\footnote{https://forums.xamarin.com/discussion/85747/xamarin-forms-feature-roadmap\#latest}
All posts are publicly available. However, for creating a new post, a user has to register in the site, which is free.

\subsection{Forums}

The Xamarin Forums platform is composed of seven general forums:
Community, General, Pre-release \& Betas, Tools and Libraries, Graphics \& Games,
Xamarin Platform and Xamarin Products.
For instance, the forum Xamarin Platform contains discussion and questions related to the development of an application for different targeted applications, while the forum Xamarin Product focuses on discussions about products related to the Xamarin technology such as TestCloud, a platform for testing the mobile app built using Xamarin.

Each forum has one or more `sub-forums'.
For instance, the mentioned Platform forum has 5 categories: Android, iOS, Cross-Platform, Mac, and Xamarin.Forms.
The forums are established in advance by the forum administrators, which means that users are not able to define new ones.

We identify two types of forums: some related to technological topics (platforms, libraries, tools, IDEs)
and others related to non-technological topics related to Xamarin (focus on event, conferences, jobs, etc.)

The main page of each forum shows a paged-list of threads (i.e., posts) and two buttons, one for creating a question and the other for creating a discussion. 
Each thread from the list shows the post's title, author, number of views, number of answers, and zero or more labels.
Those labels are colored boxes located near the title and indicate, for instance, if a question was answered, if the answered was accepted by the user who wrote the question, or if a post is an announcement.

\subsection{Posts}
A forum user can create a post for a given category.  
Even there is two types of post, i.e., questions and discussions, the creation forms of them are similar, i.e., they have the same fields.
The user can also select a list of tags associated his post. The forum suggests popular tags such as ``Android", ``async" or ``httpclient".
However, once a post is created, the forum does not show the list of tags associated to each post.\footnote{Last visit: {February, 2017}}

\subsection{Answers}
A registered user can answer an existing question or to put a `like' on existing answers.
Moreover, as in others Q\&A such as Stack Overflow, the post's author can \emph{accept} one or more answers. The accepted answers are labeled with a green-colored box.

\subsection{Users}
The forum has the functionality for registering new users, which have the right to create new posts and to write comments.
However, neither questions nor answers can be erased by a registered user once they are submitted.
In addition to a registered user, some users correspond to the Xamarin Team, which belong to the company that develops Xamarin framework, acquired by Microsoft in 2016.   


\section{Dataset construction}

The Xamarin forum does not have an API to programmatically access and retrieve the forum data, such as that one provided by Stackoverflow\footnote{https://api.stackexchange.com/} or GitHub\footnote{https://developer.github.com/v3/}.
For this reason, we developed a web crawler for retrieving and storing all data the Xamarin forum presents. 
Our engine has main two phases: Page fetching and Page parsing.

\subsection{Fetching pages}
\label{sec:fetching}
The first phase fetches (i.e., downloads) web pages written in HTML for the forum site.
The engine accesses to those pages via HTTP protocol.
We fetched two kinds of pages: 
The first kind are pages that correspond to a single post.
This pages contain post's title, question (or discussion topic), author data (names, location, roles), posting date and a list of comments. Each comment has the user name that wrote it, the date, and one or more labels such as ``Answered question".
The second kind of pages corresponds to the main pages of each forum. 
A ``main" page shows in a paged list all posts done in that forum.\footnote{In those main pages, those post are called `Threads'}
The list shows first the post corresponding to announcements, then the remaining posts ordered by decreasing creation date.
Moreover, the list shows for each post its title, the numbers of views (i.e., number of visits the post has received), numbers of comments done, and zero or more labels that indicates if the post is a question or announcement, if it has an accepted answer, etc.

\subsection{Extracting data from pages}
Then, in the second phase, our engine extracts data for each fetched page.
It parses, search and reformat data from the two kind of pages fetched.
The data we extracted is that one mentioned in the previous phase.
Then, the extracted data is stored in a  relational database.

Our engine is implemented in python language, and uses the library BeautifulSoup\footnote{https://www.crummy.com/software/BeautifulSoup/bs4/doc/} for parsing the Xamarin web pages in HTML. 
The structured data is then stored in a MySQL database using the ORM framework SQLAlchemy\footnote{http://www.sqlalchemy.org/} for connecting our Python engine app with the database.

\begin{figure}[t]

\includegraphics[trim=31cm 18cm 24cm 0cm,scale=0.6]{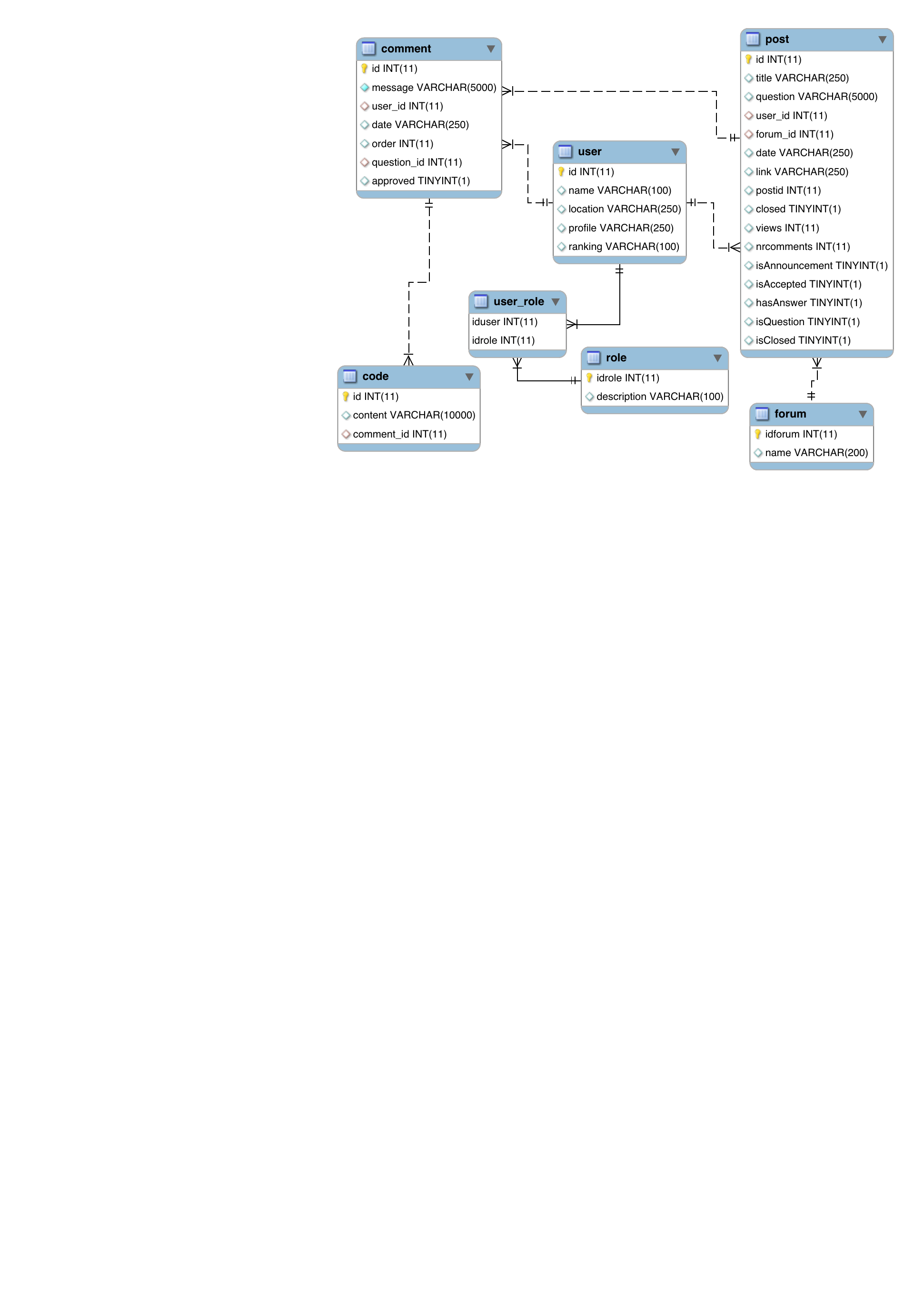}
\caption{Entity-relationship model of XamForum.}
\label{fig:diagramERR}
\vspace{-0.5cm}
\end{figure}

\subsection{Data Schema}

Figure \ref{fig:diagramERR} shows the schema of our database which stores all data we extracted from the Xamarin forum.
In this section, we explain the main entities from that schema.

The first entity is \ent{Post}, which stores all posts done in all Xamarin forums (right in the figure). Each post belongs to one \ent{Forum} and has zero or more \ent{Comments}. 
Note that the post entity stores both  questions and discussions.
We decide to model both questions and discussion in the same entity due they share the majority kind of data. 
The kind of post can be inferred from  the presence or absence of labels, which are modeled as boolean fields in table \ent{Post}. For instance, the field `isAccepted' indicates if a post has at least one comment accepted.
We decide to delegate the post classification task to the database users. 

Both \ent{Comment} and \ent{Post} entities have a relation {\it many-to-one} with the entity \ent{User}, which stores the information of the users that make or answer questions.
A \ent{User} has different \ent{Roles}. 
Registered user has the role `Member' by default, but there are other roles exclusively assigned to used belonging  the Xamarin organization such as `Forum Administrator', or `XamUProfessor" which are professionals involved on the Xamarin training program called Xamarin University.\footnote{https://www.xamarin.com/university}

Finally, the entity \ent{Code} corresponds to code chunks the users write on comments. One comment could have one or more code chunks.
We remark that this entity do not only represent programming code chunks (e.g., C\# or Java code) but also execution traces.
We let users of \xfdb to classify the data stored in \ent{Code} entity.

\section{Data Analysis}

In this section, we briefly mention the amount of the data that \xfdb stores.
\vspace{-0.2cm}
\subsection{Posts and Users}
The database has \numprint{75380} posts.
The first dates from January, 29 2013, and the last one from January, 17 2017, date we finished to carry out the data extraction from the Xamarin forum.
The number of Users that have done at least one question or one post is \numprint{32798}.
The large majority (\numprint{31310}) are users registered in the forum site, i.e., they are external to the Xamarin organization. Between the remaining users there are, for instance, 164 users from the Xamarin development team, and 41 forum administrators.

\vspace{-0.2cm}
\subsection{Comments}
The database \xfdb stores \numprint{54542} posts with at least one comment, while
\numprint{20838} posts have not received any comment, and around \numprint{15000} posts  (around 20\% of all posts) has only one comment. The maximum number of comments done in a  single post is 41.
In total, the \xfdb stores \numprint{226927} comments.

\subsection{Forums and Posts}

Table \ref{tab:forumpost} shows all Xamarin forums with the number of posts (questions or discussion) that each one has.
We observe that the majority of posts belong to ``technological" forums: almost the 30\% are related to the Xamarin.Forms technology (a Xamarin API for developing simple UI  in C\# or XAML languages), and the 23\% and 19.5\% of all posts belong to Android and iOS forums, respectively.
The fourth forum with most number of posts is about Xamarin Studio, an IDE for developing Xamarin application based on the Microsoft Visual Studio.
Between the non-technological forums, Events has \numprint{954} posts (1.27\%), where Job Listings 376 (0.5\%).
\begin{table}[htbp]
\begin{tabular}{|l|r|r|}
\hline
Forum name & {\#Posts} & {\%} \\ \hline
\hline
Xamarin Platform-Xamarin.Forms & 22478 & 29.82\% \\ \hline
Xamarin Platform-Android & 17635 & 23.39\% \\ \hline
Xamarin Platform-iOS & 14702 & 19.50\% \\ \hline
Tools and Libraries-Xamarin Studio & 3999 & 5.31\% \\ \hline
Xamarin Platform-Cross Platform & 3370 & 4.47\% \\ \hline
General & 3148 & 4.18\% \\ \hline
Tools and Libraries-Visual Studio & 2777 & 3.68\% \\ \hline
Xamarin Platform-Mac & 1708 & 2.27\% \\ \hline
Community-Events & 954 & 1.27\% \\ \hline
Xamarin Products-Xamarin Test Cloud & 880 & 1.17\% \\ \hline
Tools and Libs-Components, and Plugins & 849 & 1.13\% \\ \hline
Graphics \& Games-CocosSharp & 636 & 0.84\% \\ \hline
Xamarin Products-Xamarin Insights & 601 & 0.80\% \\ \hline
Prerelease \& Betas & 443 & 0.59\% \\ \hline
Community-Job Listings & 376 & 0.50\% \\ \hline
Community & 175 & 0.23\% \\ \hline
Tools and Libraries-Profiler & 154 & 0.20\% \\ \hline
Graphics \& Games-UrhoSharp & 152 & 0.20\% \\ \hline
Tools and Libraries-Objective Sharpie  & 123 & 0.16\% \\ \hline
Tools and Libraries-Workbooks \& Inspector & 82 & 0.11\% \\ \hline
Community-Presentations & 64 & 0.08\% \\ \hline
Xamarin Platform-Xamarin.Forms Evolution & 53 & 0.07\% \\ \hline
Others forums & 21 & $<$0.03\%\\
\hline
Total& 75380 & \multicolumn{1}{l|}{} \\ \hline
\end{tabular}
\caption{The table presents all Xamarin forums and the number of posts each forum includes.}
\label{tab:forumpost}
\vspace{-1cm}
\end{table}

\subsection{Code Chunks}
The number of total code chunks in the Xamarin forum is \numprint{73304}.
Those chunks appear in \numprint{37321} comments, that means the 16\% of comments have at least one code chunk, whereas \numprint{19673} posts have at least one chunk in any comment (representing the 36\% out of the commented posts).

\section{Research Opportunities}

In this section we present some research directions that, in our opinion,
could be done by analyzing the data from \xfdbse.


First, by analyzing discussions and questions from \xfdbse, researchers could understand the main concerns of developing cross-platform mobile applications using a state-of-the-art development framework such as Xamarin.
For instance, researchers could mine the most frequent problematic or bugs that developers face during development, 
and propose frequent bugs fix patterns mined from answers, questions and the code attached to them.
Moreover, as the code chunks from \xfdbse include execution traces, it would allow researchers to analyze commons failure during the execution of cross-applications.

Researchers could also compare efforts of developing iOS apps against developing Android apps using the Xamarin framework.

%

%
The \xfdb includes forums that discuss about tools for developing cross-platform mobile applications such as Xamarin Studio. Then, researches could  analyze the \emph{usability} of those tools from the comments written in those forums, for understanding their limitations and even propose new ones or extensions.


Even Xamarin forum is the official forum, hosted and administrated by that organization, it's not the only Q\&A site where developers post question about Xamarin platforms. 
Researcher could mine similar questions posted in two or more Q\&A sites, (e.g., Xamarin forum and Stack Overflow) and compare the answers found in those sites to, for instance, cross-validate accepted answers.


\section{Limitations}

Our fetching engine is not capable of capturing all information the Xamarin forum presents to a registered user.
For instance, a registered user can ``like" an existing answer. 
The Xamarin forum shows the number of like for each answer \emph{only} to registered user.
As our fetching engine inspects the pages to fetch as an ``anonymous user" (i.e. it is not logged to the forum site) then the number of likes is not presented.
Moreover, registered user can post images in the comments. The current version of our engine is not capable of retrieving and storing them in our database. 

\section{Related Works}


The Mining Software Repository (MSR) community has released a dump of the Q\&A site Stack Overflow \cite{MSRChallenge2013} for the mining challenge competition of 2013. 
Several works have been proposed by analyzing that dump.
For instance, Baselli et al.\cite{Bazelli2013PTS} studied the personal trails of stack overflow users.
Research have been focused on finding frequent questions for targeting a particular problematic. For example,  
Pinto et al. have mined questions to  first characterize  energy consumption problems \cite{Pinto2014MQS}. 

Several works have applied topic modeling technique such as latent Dirichlet allocation (LDA).
For example, Baura et al. \cite{Barua2014DTA} analyzed Stack Overflow data to automatically discover the main topics present in developer discussions and analyze the topic  popularity over the time.

Other works have focused on the conjunction of mining Q\&A and mobile technologies topics.
For instance, 
Linares-Vasquez et al. also used LDA to extract hot-topics from mobile-development related questions \cite{LinaresVasquez2013EAM}.
Their findings suggest that most of the questions include topics related to general questions and compatibility issues. 
Moreover, Linares-Vasquez et al.\cite{Linares-Vasquez2014ACT} studied questions and activity in Stack Overflow when  Android APIs occurs.
They found that deleting public methods from APIs is a trigger for questions that are discussed more and of major interest for the community. 
Rosen et al. \cite{Rosen2016MDA} applied (LDA) based topic models on the mobile-related Stack Overflow posts to determine what mobile developers are asking. 
Across all platforms studied, they found that questions related to app distribution,user interface, and input are among the most popular. 

Beyer et Pinzge \cite{Beyer2014} investigated 450 Android related Stack Overflow posts to get insights into the issues of Android app development. They found that the problems that are discussed most often are related to `UserInterface' and `Core Elements'. Moreover, the authors presented approach to group tag synonyms to meaningful topics\cite{Beyer2016GAT}. 
Using their approach, they reduced the number of 38,000 diverse tags on Stack Overflow to  2,481 meaningful tag.

Other works focus on analyzing developer forums.
For example, Venkatesh et al. \cite{Venkatesh2016} mined both developer forums and Stack
Overflow to find the common challenges encountered by client
developers that use Web APIs.



\section{Conclusion}

We present \xfdbse, a database that contains structured data such as discussions, questions, answers, user and their roles, extracted from the Xamarin forum, a online site where developers that use the Xamarin framework for building their cross-platform mobile  apps interact.
As developing such kind of applications is a challenge due to divergences of the mobile platforms, we envision the research community use \xfdb to better understand how is the development and maintenance of cross-platform mobile applications.  
The \xfdb is publicly available at \url{https://goo.gl/A1eh03}

\bibliographystyle{plain}
\bibliography{references}
\end{document}